\documentclass[amsmath,amssymb,aps,prx,reprint,superscriptaddress]{revtex4-2}

\usepackage{amsmath}
\usepackage{xcolor}
\usepackage{amsfonts,amssymb}
\usepackage{graphicx}
\usepackage{hyperref}
\usepackage{academicons}
\usepackage{multirow}
\usepackage{float}
\usepackage{algorithmic}
\usepackage{amsmath}

\begin{document}

\title{Crystal structure prediction with host-guided inpainting generation \\and foundation potentials}

\author{\text{Peichen Zhong}}
\email[]{zhongpc@berkeley.edu}
\affiliation{Bakar Institute of Digital Materials for the Planet, UC Berkeley, California 94720, United States}
\affiliation{Materials Sciences Division, Lawrence Berkeley National Laboratory, California 94720, United States}

\author{\text{Xinzhe Dai}}
\affiliation{Materials Sciences Division, Lawrence Berkeley National Laboratory, California 94720, United States}
\affiliation{Department of Materials Science and Engineering, UC Berkeley, California 94720, United States}

\author{\text{Bowen Deng}}
\affiliation{Materials Sciences Division, Lawrence Berkeley National Laboratory, California 94720, United States}
\affiliation{Department of Materials Science and Engineering, UC Berkeley, California 94720, United States}

\author{\text{Gerbrand Ceder}}
\affiliation{Materials Sciences Division, Lawrence Berkeley National Laboratory, California 94720, United States}
\affiliation{Department of Materials Science and Engineering, UC Berkeley, California 94720, United States}

\author{\text{Kristin A. Persson}}
\email[]{kapersson@lbl.gov}
\affiliation{Bakar Institute of Digital Materials for the Planet, UC Berkeley, California 94720, United States}
\affiliation{Materials Sciences Division, Lawrence Berkeley National Laboratory, California 94720, United States}
\affiliation{Department of Materials Science and Engineering, UC Berkeley, California 94720, United States}

\date{\today}

\begin{abstract}
Unconditional crystal structure generation with diffusion models faces challenges in identifying symmetric crystals as the unit cell size increases. We present the Crystal Host-Guided Generation (CHGGen) framework to address this challenge through conditional generation using an inpainting method, which optimizes a fraction of atomic positions within a predefined and symmetrized host structure to improve the success rate for symmetric structure generation. By integrating inpainting structure generation with a foundation potential for structure optimization, we demonstrate the method on the ZnS-P$_2$S$_5$ and Li-Si chemical systems, where the inpainting method generates a higher fraction of symmetric structures than unconditional generation. 
The practical significance of CHGGen extends to enabling the structural modification of crystal structures, particularly for systems with partial occupancy or intercalation chemistry. The inpainting method also allows for seamless integration with other generative models, providing a versatile framework for accelerating materials discovery.
\end{abstract} 

\pacs{}

\maketitle

\section{Introduction}

\begin{figure*}[t]
    \centering
    \includegraphics[width=0.9\linewidth]{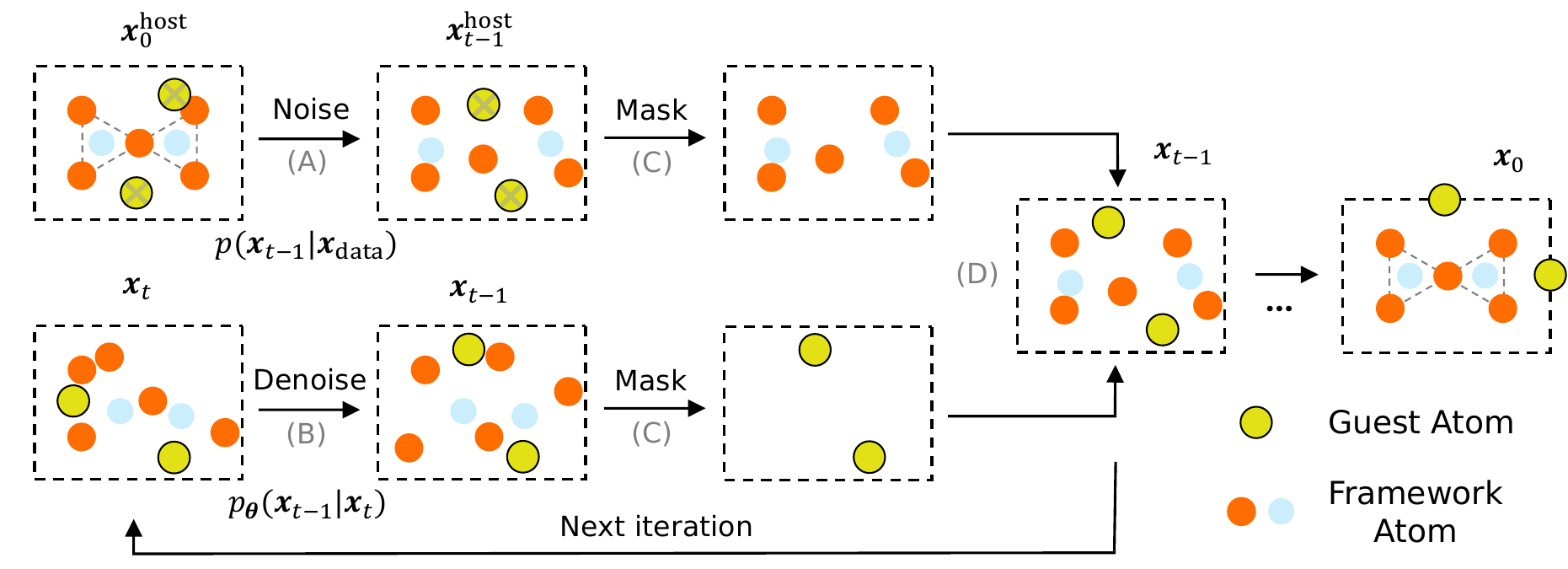}
    \caption{\textbf{Illustration of the iterative sampling strategy for structure inpainting with a host framework.} Process (A): Add noise to the host framework. Process (B): Denoise atomic configuration from $\boldsymbol{x}_{t}$ to $\boldsymbol{x}_{t-1}$ using scores $\{\boldsymbol{s}_\theta(\boldsymbol{x}_{t})\}$ predicted by the GNN. Process (C): Apply masks to the host structure and the guest atoms. Process (D): Combine the host structure and guest atoms to form the configuration $\boldsymbol{x}_t$ for the next iteration. Note that the symbol $\boldsymbol{x}^{\mathrm{host}}$ contains atomic positions of all atoms (including both host and guest atoms). The gray crosses on the guest atoms indicate that their information is not used in the processes but is retained for indexing consistency.
    }
    \label{fig:inpainting}
\end{figure*}

Crystal structure prediction (CSP) is a foundational tool in computational materials discovery with wide-ranging applications in energy storage \cite{Lu2021_cathode}, drug design \cite{Zhou2025_drug}, and superconductors \cite{Zhang2022_superconductor}. The ability to predict stable atomic arrangements for a given chemical composition is critical for materials design, yet remains a challenge due to the high dimensionality of chemical and configurational space \cite{Gusev2023_optimal}. Traditional computational approaches using density functional theory (DFT) calculations have achieved notable successes, such as random structure searches \cite{pickard2011ab}, genetic algorithms \cite{oganov2006crystal}, particle-swarm optimization \cite{PhysRevB.82.094116_PSO}, substitution models \cite{Hautier2011a}, and exact lattice model approaches \cite{Huang2016_MAXSAT}. 
However, DFT-based search algorithms can become computationally prohibitive, especially when applied to multi-component systems with complex compositional spaces \cite{Ferrari2023_SRO_review}. 

Recent advances in graph neural network (GNN)-based machine learning models have introduced promising alternatives to traditional CSP methods, with a key milestone being the development of foundation potential, or universal machine learning interatomic potentials---offering accurate and transferable modeling across diverse material systems \cite{chen2022_m3gnet, Deng2023_chgnet, batatia2023foundation, Kim2025_sevenet}. These foundation potentials trained on millions of DFT calculations demonstrate remarkable generalizability in exploring vast chemical spaces for materials discovery \cite{merchant2023scaling, zhang2024dpa, yang2024mattersim, fu2025learning}.
Another emerging direction is deep-generative models, particularly diffusion models, which learn the data manifold or probabilistic distribution and generate new configurations via stochastic or variational approaches \cite{Ren2022_VAE, park2024exploration, Cheng2024_response}. \citet{xie2021_cdvae} introduced CDVAE that uses a variational autoencoder to sample lattice parameters and compositions and a diffusion model to optimize atomic coordinates. 
Although promising, CDVAE-generated structures are predominantly thermodynamically unstable or lack symmetry \cite{szymanski2025establishing, levy2025symmcd}. 
\citet{kurz_efficient_2014} introduced a wrapped normal distribution to effectively couple lattice diffusion with fractional coordinates, a strategy that was successfully implemented in DiffCSP. \cite{kurz_efficient_2014, bortoli_riemannian_2022, jiao_crystal_2023}. \citet{Zeni2025_mattergen} further adapted the scheme with edge features for lattice scores in the MatterGen framework that enables diffusions on lattices, fractional coordinates, and chemical species. By learning from materials datasets such as Materials Project (MP) \cite{Jain2013} and Alexandria \cite{Schmidt2023_Alexandria}, 
MatterGen is capable of conducting scalable and universal exploration in high-dimensional design space and achieves excellent performance in structural stability, uniqueness, and novelty, albeit with an observed limitation to a smaller scale (e.g., $N_\text{atom}\leq 20$). 
Beyond GNN-based models, other approaches such as U-net-based diffusion models \cite{yang2023scalable}, optimization of subcell structures from amorphous configurations \cite{Aykol2024_a2c}, and large language models \cite{gruver2024_LLM, Antunes2024_LLM} have shown promise in crystal structure generations without the need to limit structure sizes.

In this work, we extend GNN-based diffusion models to enable fractional crystal structure design via inpainting generation---that is, given a host or substrate structure, we optimize the placement of additional 'guest' atoms within the existing framework. This application is particularly valuable in several material domains, e.g., defective materials, intercalation electrodes \cite{Li2024_Lifree}, molecular absorption on catalyst surfaces \cite{liao2023equiformerv2}, and interfacial solid reactions where surfaces reconstruct while bulk structures remain unchanged \cite{Ronne2024_surface}. We first summarize the fundamentals of diffusion models and inpainting generation, and discuss the locality bias of GNN-based diffusion models, particularly as a key challenge when generation is performed at large scales. To address these gaps, we introduce \textbf{C}rystal \textbf{H}ost-\textbf{G}uided Generation (CHGGen), which integrates inpainting generation based on symmetrized frameworks and foundation potential for structure optimization. We demonstrate the effectiveness of host-guided generation through a case study on CSP within the ZnS–P$_2$S$_5$ chemical space, and showcase the broader applicability of CHGGen across the continuous chemical space of the Li$_x$–Si alloy system. Finally, we discuss the limitations and potential opportunities of applying CHGGen with state-of-the-art generative models for CSP problems in future directions.

\section{Theory}
We first briefly recap the concepts of diffusion models with score-based denoising and then introduce inpainting as a conditional generation method for the structural modification of crystal structures.

\noindent\textbf{Diffusion model.}
Generating samples from a probability density function $p(\boldsymbol{x})$ in high-dimensional space $\mathbb{R}^d$ can be achieved by modeling the gradient of the log-probability density, known as the \emph{score function} $\nabla_{\boldsymbol{x}} \log p(\boldsymbol{x})$ in diffusion models.
\citet{song2020score} demonstrated that both the diffusion process and its reverse can be formulated as stochastic differential equations (SDE)
\begin{equation}\label{eq:forward_SDE}
    \mathrm{d} \boldsymbol{x} = \boldsymbol{f}(\boldsymbol{x}, t) \mathrm{d}t + g(t) \mathrm{d}\boldsymbol{w},
\end{equation}
\begin{equation}\label{eq:reverse_SDE}
    \mathrm{d} \boldsymbol{x} = \left[\boldsymbol{f}(\boldsymbol{x}, t) - g^2(t) \nabla_{\boldsymbol{x}} \log p_t (\boldsymbol{x})\right] \mathrm{d}t + g(t) \mathrm{d}\boldsymbol{\overline{w}},
\end{equation}
where $\boldsymbol{w}$ and $\boldsymbol{\overline{w}}$ represent the standard Brownian motion process and its time-reversed analogue, respectively.  $\boldsymbol{f}(\boldsymbol{x},t)$ is the drift coefficient and $g(t)$ is the diffusion coefficient of $\boldsymbol{x}(t)$. $p_t(\boldsymbol{x})$ denotes the probability density of $\boldsymbol{x}(t)$. Here $t$ is the time variable $t\in [0,T]$ to describe the diffusion process $\{\boldsymbol{x}(t)\}_{t=0}^{T}$.

Eq.~\eqref{eq:forward_SDE} describes the forward process to corrupt the data distribution $\boldsymbol{x}(0) \sim  p_0(\boldsymbol{x})$ to obtain the prior distribution $\boldsymbol{x}(T) \sim  p_T(\boldsymbol{x})$, which follows a uniform distribution. Eq.~\eqref {eq:reverse_SDE} describes the reverse process to sample $\boldsymbol{x}(0)$ by solving the reverse SDE with the score term $\nabla_{\boldsymbol{x}} \log p(\boldsymbol{x})$. For crystal structure generation, we adopt the variance-exploding (VE) diffusion scheme for the atomic coordinates, where the process $\{\boldsymbol{x}(t)\}_{t=0}^T$ is given by the SDE
\begin{equation}
    \mathrm{d} \boldsymbol{x} = \sqrt{\frac{\mathrm{d} [\sigma^2(t)]}{\mathrm{d} t}} \mathrm{d} \boldsymbol{w}.
\label{eq:VE-SDE}
\end{equation}
Here $\{\sigma(t)\}$ is a sequence of exponentially increasing standard deviations given $\sigma_{\min} = \sigma_1, \cdots, \sigma_T=\sigma_{\max}$. The VE-SDE is particularly suitable for atomic coordinates in crystals, as VE-SDE does not induce disconnected graphs at the large noisy limit under periodic boundary conditions.

The samples can be generated using ancestral sampling, where successive states are sampled according to:
\begin{equation}
    \boldsymbol{x}_{t-1} = \boldsymbol{x}_t + (\sigma_t^2 - \sigma_{t-1}^2) \boldsymbol{s}_{\boldsymbol{\theta^*}}(\boldsymbol{x}_t, t) + \boldsymbol{z} \sqrt{\frac{\sigma_{t-1}^2(\sigma_t^2 - \sigma_{t-1}^2)}{\sigma_t^2}}
    \label{eq:discretized_reverse_SDE}
\end{equation}
where $\boldsymbol{x}_T \sim \mathcal{N}(\mathbf{0}, \sigma_T^2\mathbf{I})$, and $\boldsymbol{z} \sim \mathcal{N}(\mathbf{0}, \mathbf{I})$. In the continuous limit, $\sqrt{\sigma_{t-1}^2/\sigma_t^2}\approx 1$. The implementation is achieved using a predictor-corrector sampling strategy with the Langevin corrector. We refer the readers to Ref.~\cite{song2020score} for mathematical details of score-based SDE and sampling strategies.

\begin{figure}[htb]
\begin{algorithmic}
        \STATE \hrulefill
        \STATE {\bfseries Unconditional Generation.}~Inputs: Randomly initialized atomic positions $\boldsymbol{x}_{T}$. Signal-to-noise ratio $\delta$. Number of predictor steps $T$; Number of corrector steps $M$.
        \STATE \hrulefill
        \FOR{$ t = T, \cdots, 1$}
        \STATE $\boldsymbol{x}_{t-1} \leftarrow \boldsymbol{x}_{t} + (\sigma_{t}^2 -\sigma_{t-1}^2)\boldsymbol{s}_{\theta}(\boldsymbol{x}_{t}, t) $
        \STATE $\boldsymbol{z}\sim \mathcal{N}(\mathbf{0}, \mathbf{I})$
        \STATE $\boldsymbol{x}_{t-1} \leftarrow \boldsymbol{x}_{t-1} + \sqrt{\frac{\sigma_{t-1}^2(\sigma_{t}^2 -\sigma_{t-1}^2)}{\sigma_{t}^2}}\ \boldsymbol{z}$
        \FOR{$j = 1, \cdots, M$}
        \STATE $\boldsymbol{z}\sim \mathcal{N}(\boldsymbol{0}, \boldsymbol{1})$
        \STATE $\boldsymbol{g} \leftarrow \boldsymbol{s}_{\theta}(\boldsymbol{x}_{t-1}, t-1)$
        \STATE $\epsilon \leftarrow 2(\sqrt{3} \delta / \Vert \boldsymbol{g} \Vert_2)^2$
        \STATE $\boldsymbol{x}_{t-1} \leftarrow \boldsymbol{x}_{t-1} + \epsilon \ \boldsymbol{g} +\sqrt{2\epsilon} \ \boldsymbol{z}$
        \ENDFOR
        \ENDFOR
        \STATE \hrulefill
    \end{algorithmic}
\end{figure}

\noindent\textbf{Denoising score matching.} 
To estimate the score function $\nabla_{\boldsymbol{x}} \log p_t (\boldsymbol{x})$, we use score matching (SM) to optimize the model parameters $\theta^*$ by minimizing
\begin{equation}
    \begin{aligned}
        \mathcal{L}_{\mathrm{SM}} &= \mathbb{E}_{p_t(\boldsymbol{x})} \left[ \left\Vert \boldsymbol{s}_\theta(\boldsymbol{x}(t), t) - \nabla_{\boldsymbol{x}} \log p_t(\boldsymbol{x}) \right\Vert^2 \right].
    \end{aligned}
    \label{eq:score_matching}
\end{equation}
$\mathbb{E}_{p_t(\boldsymbol{x})}\left[\cdot\right]$ represents the expectation value with respect to the probability distribution $p_t(\boldsymbol{x})$, which can be approximated by a Gaussian transition probability $p(\boldsymbol{x}(t) | \boldsymbol{x}(0)) \propto e^{-[\boldsymbol{x}(t) - \boldsymbol{x}(0)]^2/2\sigma^2}$ such that Eq.~\eqref{eq:score_matching} is formulated as denoising score matching (DSM) with \cite{vincent_connection_2011, ho2020denoising}
\begin{equation}
    \begin{aligned}
        \mathcal{L}_{\mathrm{DSM}} &= \mathbb{E}_{\boldsymbol{x}(0)} \mathbb{E}_{\boldsymbol{x}(t) | \boldsymbol{x}(0)} \left[ \left\Vert \boldsymbol{s}_\theta(\boldsymbol{x}(t), t) + \frac{\boldsymbol{\epsilon}}{\sigma} \right\Vert^2 \right].
    \end{aligned}
    \label{eq:denoise_score_matching}
\end{equation}
Here $\boldsymbol{x}(0) \sim p_0(\boldsymbol{x})$ and $\boldsymbol{x}(t) \sim p(\boldsymbol{x}(t) | \boldsymbol{x}(0))$, $s_\theta(\boldsymbol{x}(t), t)$ is the score function predicted by the GNN model, $\boldsymbol{\epsilon}$ represents the normalized noise $[\boldsymbol{x}(t) - \boldsymbol{x}(0)]/\sigma \sim \mathcal{N}(\mathbf{0},\mathbf{I})$. In the training process, $\sigma$ is sampled uniformly from the interval $[\sigma_{\min}, \sigma_{\max}]$ to perturb the configuration $\boldsymbol{x}(0)$ and obtain the noisy configuration $\boldsymbol{x}(t)$ to construct the DSM loss in Eq.~\eqref{eq:denoise_score_matching}.

\noindent\textbf{Inpainting.}
Inpainting is a \textit{conditional} generation process where a model completes missing elements within a given context. Inpainting has demonstrated significant applications in materials and chemistry, including the discovery of chemical reaction transition states \cite{duan2023accurate} and the generation of symmetry-constrained 2D materials \cite{li2024_inpainting_2D}. In CSP, inpainting enables the optimal placement of additional atoms (termed guest atoms) within a predefined host crystal structure, where a binary masking strategy applies different noise treatments to known regions (host structure) and unknown regions (areas to be inpainted -- guest atoms).

Unlike training a certain distribution of the mask, \citet{lugmayr_repaint_2022} introduced the \texttt{repaint} algorithm (inpainting + resampling) for high-quality 2D image inpainting using diffusion models. One can simply train the diffusion model with DSM to learn the joint distribution. During inference, the conditional distribution is approached using the resampling technique for inpainting generation. As shown in the Algorithm, in addition to the unconditional generation steps, the resampling repeatedly "jumps back" in the diffusion process and resamples the unknown regions multiple times with $r$ steps) at each timestep $t$, with a mask $m$ to separate the host and guest atoms. This resampling procedure helps harmonize the generated content with existing regions by allowing multiple attempts at generating coherent inpainted content. For detailed implementation and theoretical foundations, we refer readers to Ref.~\cite{lugmayr_repaint_2022} and \cite{dai2024inpainting} for details of this approach.

\begin{figure}[!h]
\begin{algorithmic}
        \STATE \hrulefill
        \STATE {\bfseries Inpainting Generation.}~Inputs: Atomic positions of unperturbed host structure with randomly initialized guest atoms $\boldsymbol{x}^{\mathrm{host}}_0$; Atomic positions of all atoms sampled randomly in the unit cell $\boldsymbol{x}_{T}$; Mask for guest atoms $m$; Signal-to-noise ratio $\delta$; Number of predictor steps $T$; Number of corrector steps $M$; Number of resampling steps $r$.
        \STATE \hrulefill
        \FOR{$ t = T, \cdots, 1$}
        \FOR{$n = 1, \cdots, r$}
        \STATE $\boldsymbol{x}_{t-1} \leftarrow \boldsymbol{x}_{t} + (\sigma_{t}^2 -\sigma_{t-1}^2)\boldsymbol{s}_{\theta}(\boldsymbol{x}_{t}, t) $
        \STATE $\boldsymbol{z}\sim \mathcal{N}(\mathbf{0}, \mathbf{I})$
        \STATE $\boldsymbol{x}_{t-1} \leftarrow \boldsymbol{x}_{t-1} + \sqrt{\frac{\sigma_{t-1}^2(\sigma_{t}^2 -\sigma_{t-1}^2)}{\sigma_{t}^2}}\ \boldsymbol{z}$
        \FOR{$j = 1, \cdots, M$}
        \STATE $\boldsymbol{z}\sim \mathcal{N}(\boldsymbol{0}, \boldsymbol{1})$
        \STATE $\boldsymbol{g} \leftarrow \boldsymbol{s}_{\theta}(\boldsymbol{x}_{t-1}, t-1)$
        \STATE $\epsilon \leftarrow 2(\sqrt{3} \delta / \Vert \boldsymbol{g} \Vert_2)^2$
        \STATE $\boldsymbol{x}_{t-1}  \leftarrow \boldsymbol{x}_{t-1}  + \epsilon \ \boldsymbol{g} +\sqrt{2\epsilon} \ \boldsymbol{z}$
        \ENDFOR
        \STATE $\boldsymbol{x}_{t-1}^{\mathrm{host}} \leftarrow \boldsymbol{x}^{\mathrm{host}}_0 + \sigma_{t-1} \boldsymbol{z}$
        \STATE $\boldsymbol{x}_{t-1} \leftarrow (1-m) \odot \boldsymbol{x}^{\mathrm{host}}_{t-1} + m \odot \boldsymbol{x}_{t-1} $
        \IF{$n < r$ \AND $t>1$}
        \STATE $\boldsymbol{z}\sim \mathcal{N}(\boldsymbol{0}, \boldsymbol{1})$
        \STATE $\boldsymbol{x}_{t} \leftarrow \boldsymbol{x}_{t-1} + \sqrt{\sigma_{t-1}^2 - \sigma_{t-2}^2}\ \boldsymbol{z}$
        \ENDIF
        \ENDFOR
        \ENDFOR
        \STATE \hrulefill
    \end{algorithmic}
    \label{alg:inpainting}
\end{figure}

\begin{figure*}[htb]
    \centering
    \includegraphics[width=0.9\linewidth]{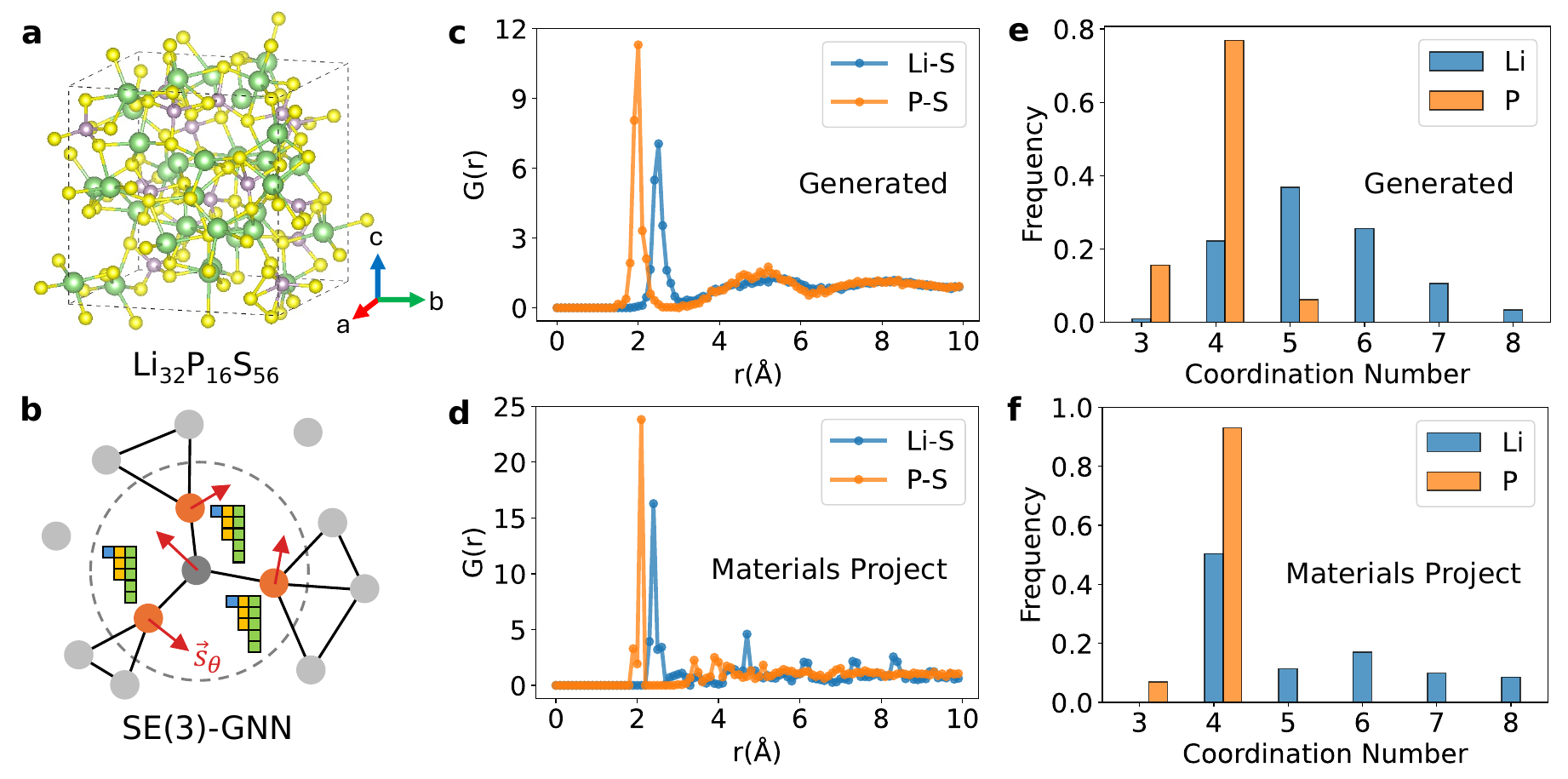}
    \caption{\textbf{Analysis of the locality bias in GNN-based diffusion models.} (a) A generated supercell structure of Li$_{32}$P$_{16}$S$_{56}$ exhibiting an amorphous configuration. (b) Illustration of SE(3)-equivariant graph neural networks, where the score function $\{\boldsymbol{s}_\theta(\boldsymbol{x}_{t})\}$ is predicted as a vector from each graph node (red arrows). (c--d) Radial distribution functions (RDF) of Li--S and P--S in generated structures compared with database structures from the MP. (e--f) Comparison of local chemical environments grouped by coordination number between generated structures and MP structures.}
    \label{fig:LiPS}
\end{figure*}

Figure \ref{fig:inpainting} illustrates the iterative sampling procedure for inpainting generation. During each reverse diffusion step, the process follows several distinct stages:
First, the atoms in the host structure are perturbed using Gaussian noise determined by the noise scheduler of the subsequent step $\sigma_{t-1}$ (Process A):
\begin{equation}
    \boldsymbol{x}_{t-1}^{\mathrm{host}} = \boldsymbol{x}_{0}^{\mathrm{host}} + \sigma_{t-1} \boldsymbol{z}, \quad \boldsymbol{z} \sim \mathcal{N}(\mathbf{0}, \mathbf{I}).
\end{equation}
This ensures that both the guest atoms and host structure in the current configuration $\boldsymbol{x}_t$ maintain comparable noise scales. The GNN model then computes the score using all atoms and executes the reverse diffusion via Eq.~\eqref{eq:discretized_reverse_SDE} (Process B). In our notation, $\boldsymbol{x}^{\mathrm{host}}$ contains atomic positions of all atoms (including the framework and guest atoms) for indexing consistency. We apply masks $(1-m)$ to $\boldsymbol{x}^{\mathrm{host}}$ and $m$ to $\boldsymbol{x}$ (Process C) to construct $\boldsymbol{x}_{t-1}$ (Process D):
\begin{equation}
    \boldsymbol{x}_{t-1} = (1-m) \odot \boldsymbol{x}_{t-1}^{\mathrm{host}} + m \odot \boldsymbol{x}_{t-1}
\end{equation}
Through this iterative reverse diffusion process, the noise scale $\{\sigma_t\}$ gradually decreases, resulting in a final crystal structure that is closely aligned with the original host structure with minimal deviation (e.g., $\sigma_{\text{min}} = 0.001$ \AA). The positions of the guest atoms are therefore determined by the distribution conditioned on the host structure.


\section{Results}
We developed an SE(3)-equivariant Graph Neural Network (GNN) based on the \verb|NequIP| architecture to predict the score function for both unconditional and inpainting diffusion processes. The model training followed the DSM scheme described in Eq.~\eqref{eq:denoise_score_matching}, where the dataset was prepared using crystal structures from the MP database with energy above hull $E_{\text{hull}} < 0.1$ eV (see Methods in SI).

In the following sections, we first examine the locality bias encountered when generating structures with large unit cells through unconditional generation. These insights led to the development of the CHGGen framework. We demonstrate the effectiveness of CHGGen on two example chemical systems: Zn-P-S and Li-Si, which are complemented by CHGNet as a foundation potential for iterative structural optimizations and thermodynamic stability screening. Additionally, we also demonstrate example studies of CSP of 16 compositions used in DiffCSP \cite{jiao_crystal_2023} and solid-solid interface in Supplementary Information.

\begin{figure*}[t]
    \centering
    \includegraphics[width=0.9\linewidth]{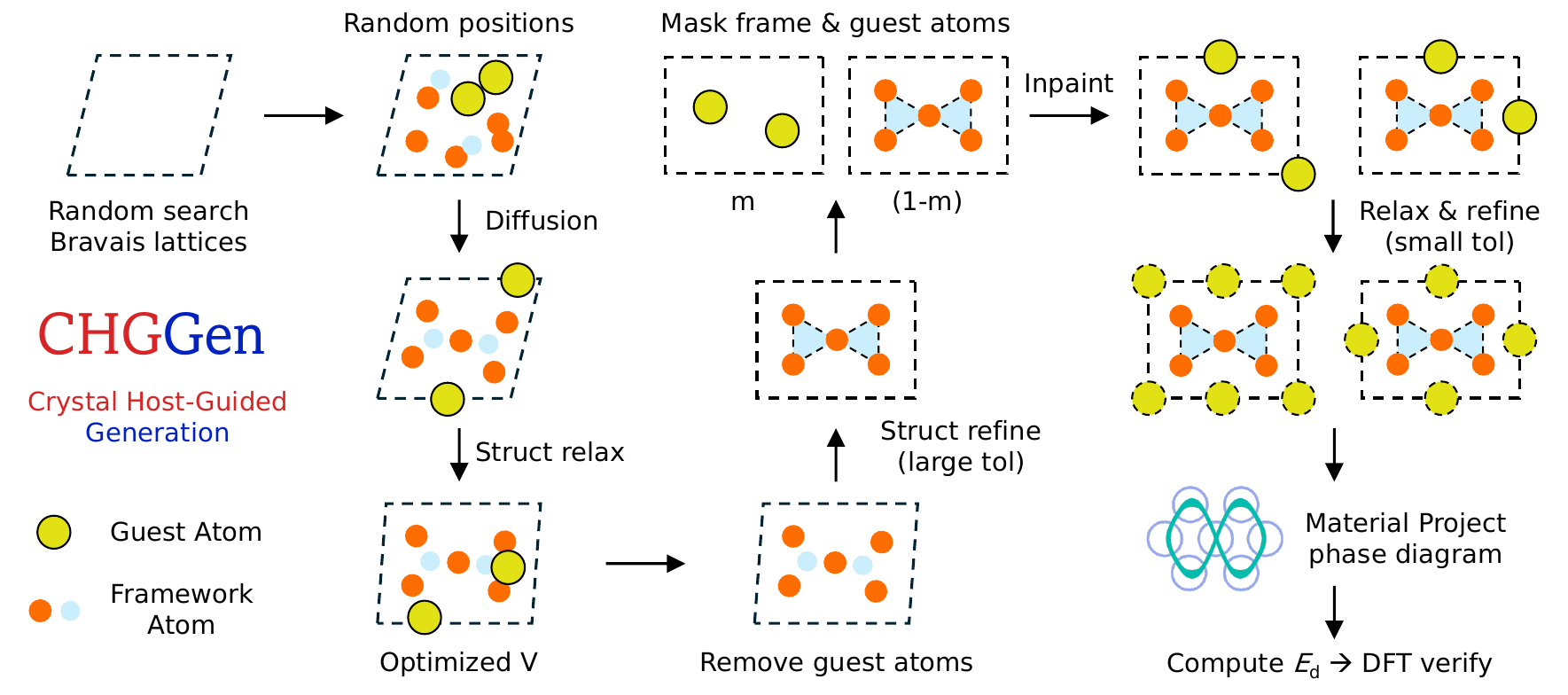}
    \caption{\textbf{Computational workflow of CHGGen.} The process begins with a random search for Bravais lattices containing a specified number of atoms, followed by an unconditional generation with reverse diffusion and structure relaxation using CHGNet. Structure refinement is applied after removing guest atoms to obtain a symmetrized framework. Inpainting generation is then performed based on this refined framework to guide the creation of complete crystal structures. Finally, the generated structures undergo relaxation to determine decomposition energy, with promising candidates (those exhibiting low decomposition energy) selected for DFT verification. The dashed circles represent crystallographically equivalent atomic positions in a crystal structure.}
    \label{fig:chggen_framework}
\end{figure*}

\subsection{Locality bias of GNN-based generative models}

To investigate the limitations of unconditional generation at large scales, we generated 10 supercell configurations of (Li$_2$S-P$_2$S$_5$)$_{10}$ and (Li$_4$S$_2$-P$_2$S$_5$)$_8$ with the parametrized diffusion models (Fig.~\ref{fig:LiPS}a).
Figure~\ref{fig:LiPS}a shows a snapshot of the generated structures, where the long-range periodicity is absent and an amorphous configuration is exhibited.
Figure~\ref{fig:LiPS}c and \ref{fig:LiPS}d present the radial distribution functions (RDF) of the generated structures and the structures from the MP database, showing similar major peaks in both RDF plots. The RDF suggests that the generated structures exhibit physically reasonable Li–S and P–S bonding environments, which represent the learned local distribution from the dataset.

We evaluated local coordination environments using the \texttt{LocalGeometryFinder} toolkit \cite{Waroquiers2017_statistical_analysis} and classified the local environments based on coordination numbers. In Fig.~\ref{fig:LiPS}e, P atoms predominantly occupy tetrahedral sites (4-coordinated), consistent with known Li–P–S crystal structures \cite{Lee2023_LiPS}. In contrast, Li exhibits a broad distribution of coordination numbers, with peaks at 5-fold ($\sim$40\%), 4-fold ($\sim$25\%), and 6-fold ($\sim$30\%) geometries. 
Notably, these coordination statistics qualitatively align with patterns observed in the MP training dataset (Fig.~\ref{fig:LiPS}f), where P atoms maintain rigid tetrahedral coordination while Li atoms display more variable environments.

Based on the successful learning and reconstruction of local distribution from the generative model, we hypothesize that the failure to propagate long-range order in generated structures stems from two interrelated factors: (1) Locality bias in GNNs: While the model effectively captures short- to medium-range atomic correlations, its finite receptive field constrains the learning of global crystallographic patterns.
(2) Stochasticity in reverse diffusion: The stochastic differential equation for reverse diffusion processes inherently samples from a learned distribution of the entire dataset.
Without coupling the atomic arrangements and supercell parameters, the diffusion process tends to sample from the entire distribution of the dataset in the generated structures, rather than from a narrowed distribution in specific crystal systems. Consequently, structures with large unit cells manifest as "mosaics" of local structure motifs rather than coherent crystalline structures.
These limitations may be universal even with lattice diffusion, as the GNN architecture lacks explicit mechanisms to maintain long-range crystallographic order when featuring the atomic configurational space \cite{Gong2023}.

\subsection{Crystal host-guided generation}

\begin{figure*}[hbt]
    \centering
    \includegraphics[width=0.95\linewidth]{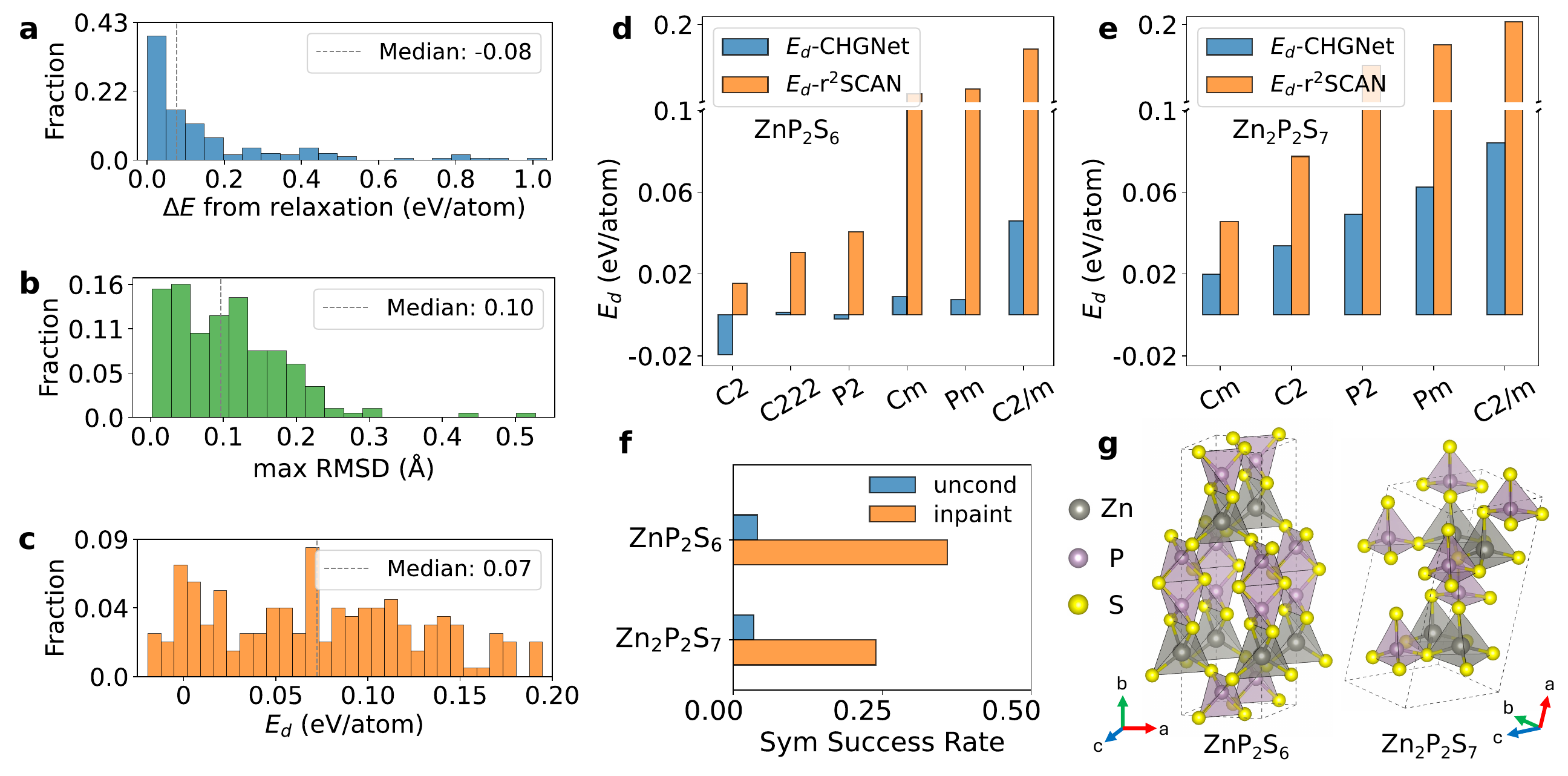}
    \caption{\textbf{Generation results in the Zn-P-S system demonstrating local stability, global stability, and capability for identifying symmetric structures.} (a) Energy change ($\Delta E$) following structure relaxation. (b) Maximum pair-wise root-mean-squared displacement (RMSD) representing differences between generated and relaxed structures. (c) Distribution of decomposition energies ($E_d$) of generated structures relative to the MP phase diagram as predicted by CHGNet. (d--e) Comparison of decomposition energies predicted by CHGNet (blue) and r$^2$SCAN-DFT (orange) relative to the MP phase diagram at GGA/GGA$+U$ (blue) and r$^2$SCAN (orange) levels of accuracy. (f) Success rate of identifying symmetric crystal structures using unconditional vs. inpainting generation. (g) Examples of generated structures with the lowest DFT decomposition energy in ZnP$_2$S$_6$ (C2, $E_d$ = 0.015 eV/atom) and Zn$_2$P$_2$S$_7$ (Cm, $E_d$ = 0.046 eV/atom)}.
    \label{fig:ZnPS}
\end{figure*}

Motivated by our observation that GNN-based diffusion models struggle to generate crystalline structures with long-range periodic structures, we developed the CHGGen framework as a targeted approach to mitigate this limitation. CHGGen integrates three key components: (1) unconditional structure generation, (2) inpainting generation based on symmetry-refined host structure, and (3) structural optimization using the CHGNet.

Figure~\ref{fig:chggen_framework} illustrates the CHGGen computational workflow. The process begins with sampling various Bravais lattices at a fixed volume through a random search over lattice constant ratios and angles (see Methods in SI). The unit cell volume is determined as $N*V_0$, where $N$ represents the number of atoms and $V_0$ denotes the atomic volume. The $V_0$ can be initialized either from related crystalline phases or predicted by composition-based regression models \cite{Goodall2020_roost}. This atomic volume serves as prior information subject to optimization in subsequent steps.
Following lattice determination, fractional coordinates for all atoms are initialized with random numbers drawn from $\mathcal{N}(0,1)$. The diffusion process then proceeds by solving the reverse SDE using scores predicted by the SE(3)-GNN (unconditional generation). Given that the volumes and lattices of the generated structures are drawn from simple priors and random search, the CHGNet is employed for structure relaxation to optimize both unit cells and atomic coordinates. This process represents a well-defined local energy minima search task and does not suffer from the locality bias encountered during the diffusion process.

The next phase is initiated by removing atoms that exhibit broad local environment distributions (e.g., Li). The remaining structure (framework) undergoes symmetry refinement using \texttt{spglib} through incremental structural matching tolerance to obtain a space group with higher symmetry (i.e., until the space group is not P1). 
Since the guest atoms exhibit diverse local environment distributions, refinement without them is more feasible for obtaining the symmetric structure. The fractional coordinates of the removed guest atoms are then reinitialized from $\mathcal{N}(\mathbf{0},\mathbf{I})$ within the symmetrized framework, and inpainting generation is performed using masks $m$ and $(1-m)$ for the guest and framework atoms, respectively.

The inpainting-generated structures are further relaxed using CHGNet and structure refinement is performed with a small tolerance to obtain the spacegroup. The CHGNet-calculated energy for the relaxed structure is used to determine the decomposition energy $E_d$ relative to the MP phase diagram at the GGA/GGA+$U$ level of accuracy. Finally, structures with $E_d$ within a specified threshold (e.g., $E_d < 0.1$ eV/atom) are submitted for DFT calculations to obtain more accurate thermodynamic stability assessments. In our studies, we used the r$^2$SCAN functional to evaluate the DFT decomposition energy against the MP r$^2$SCAN phase diagram.

\begin{figure*}[htb]
    \centering
    \includegraphics[width=0.9\linewidth]{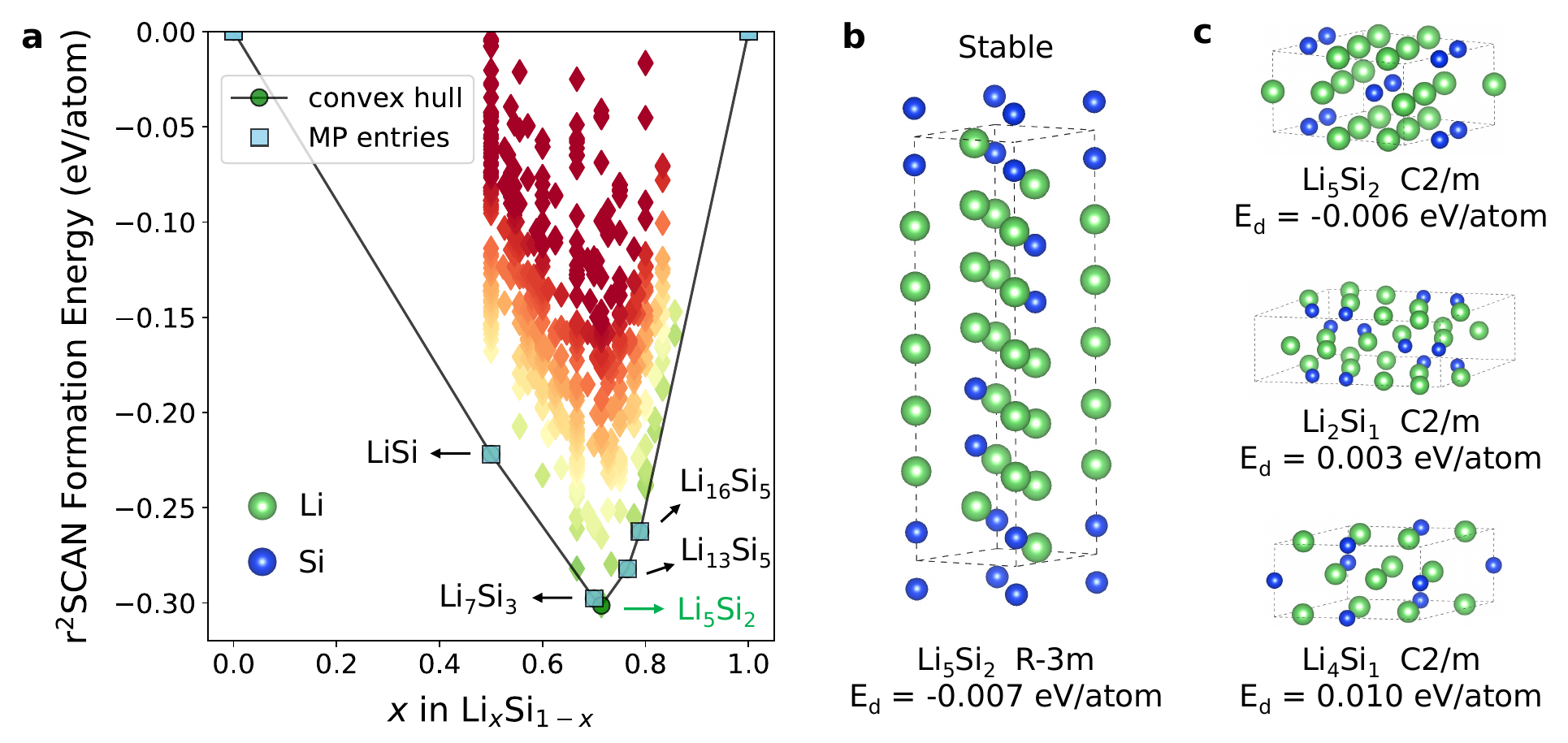}
    \caption{\textbf{Formation energies and crystal structures in the Li-Si chemical system.} (a) Formation energy phase diagram calculated with generated structures in the Li-Si chemical system using r$^2$SCAN-DFT. Blue squares represent stable structures from the MP database, green dots indicate stable compounds on the formation energy convex hull, and the diamonds are the metastable compounds above the convex hull.  (b) Stable generated structure (Li$_5$Si$_2$, R-3m) confirmed by DFT calculations ($E_d = -0.007$ eV/atom). (c) Generated metastable structures with C2/m spacegroup in Li$_5$Si$_2$, Li$_2$Si, and Li$_4$Si compositions, respectively.}
    \label{fig:LiSi}
\end{figure*}

\subsection{Example: Zn-P-S}

The first example predicts the crystal structure in the ZnS-P$_2$S$_5$ chemical space, which represents a logical extension of the related Li-P-S system that exhibits various stable and metastable polymorphs along the Li$_2$S-P$_2$S$_5$ compositional line \cite{Szczuka2022_LiPS}. Understanding phase stability in analogous Zn-based systems is important for advancing Zn-based solid-state batteries.

We focused on the CSP of ZnP$_2$S$_6$ and Zn$_2$P$_2$S$_7$ using CHGGen. To assess the local stability of the generated structures, we evaluated the structural and energetic differences between the initially generated structures and their CHGNet-relaxed counterparts.
In Fig.~\ref{fig:ZnPS}a and \ref{fig:ZnPS}b, we present the energy and geometrical differences between the relaxed structures and generated structures. Most of the generated structures exhibit energy changes of $\Delta E < 0.1$ eV/atom, with a median value of 0.08 eV/atom. The geometric differences, quantified by maximum pair-wise root-mean-squared distance (RMSD), show a median value of 0.10 \AA\ between relaxed and generated structures. As illustrated in Fig.~\ref{fig:ZnPS}a-b, outliers with large energy changes correspond to RMSD values exceeding 0.3 \AA, indicating that most of the generated structures are close to the local minima and can be reasonably searched using foundation potential structure relaxation.

Figure~\ref{fig:ZnPS}c presents the CHGNet-predicted decomposition energies ($E_d$) to quantify the thermodynamic stability. The distribution shows a median $E_d$ of 0.07 eV/atom, indicating that the majority of generated structures are metastable in this chemical space, while 6.5\% of the generated structures are identified as stable with $E_d<0$ at the CHGNet level of accuracy.

After r$^2$SCAN-DFT calculations and reevaluation of $E_d$ with respect to the MP-r$^2$SCAN phase diagram, the structure with negative $E_d$ was confirmed to be metastable. This discrepancy can be attributed to CHGNet's approximate 30 meV/atom prediction error, which may alter the predicted phase stability when $E_d$ values are small. As illustrated in Fig.~\ref{fig:ZnPS}d and \ref{fig:ZnPS}e, while CHGNet demonstrates good overall agreement with DFT, it systematically underestimates the decomposition energy for structures with high values ($E_d > 0.1$ eV/atom), consistent with the known softening effect of foundation potentials \cite{Deng_softening_2025}. This finding suggests that practitioners could consider using a lower threshold (e.g., $E_d < 30$ meV/atom) when screening generated structures for DFT validation.

To evaluate how diffusion models perform in generating symmetric crystal structures, we define a simple metric \emph{symmetry success rate} as the fraction of generated structures possessing space groups with higher symmetry than P1 or P-1 after relaxation and refinement (see Methods in SI).
Figure~\ref{fig:ZnPS}f compares the symmetry success rates from unconditional (blue) and inpainting (orange) generations.
The inpainting approach uses the refined P--S frameworks to achieve significantly higher success rates compared to the unconditional ones ($<5\%$), which highlights the effectiveness and practical advantages of conditional generation with structural priors. 
The structures with the lowest-$E_d$ for ZnP$_2$S$_6$ and Zn$_2$P$_2$S$_7$ are illustrated in Fig.~\ref{fig:ZnPS}g. While the generated structures are predicted to be metastable by DFT calculations, this example demonstrates CHGGen's potential for exploring crystal structures in the chemical space that is currently absent from existing databases.

\subsection{Example: Li-Si alloys}

The second example extends to a binary system -- Li-Si alloys. The Li-Si system has particular significance as a high-capacity anode in Li-ion batteries\cite{Tan2021_Si_science}. The MP database contains 13 DFT-calculated structures, of which 4 are thermodynamically stable ($E_{\text{hull}} = 0$) at zero~K.

To demonstrate the effectiveness of identifying low-energy polymorphs, we performed structure generation for Li-rich phases (Li$_x$Si$_y$, where $x>y$, a composition range known to contain many stable phases \cite{Artrith2018_LiSi}) following the workflow outlined in Fig.~\ref{fig:chggen_framework} (see Methods in SI).
After CHGGen generation and r$^2$SCAN-DFT calculations, we constructed the phase diagram using DFT formation energies from both MP structures and the generated structures.  In Fig.~\ref{fig:LiSi}a, green circles represent new stable polymorphs identified by CHGGen, while blue squares indicate the reported stable MP structures. We calculated decomposition energies using the MP phase diagram (without the on-the-hull structure of Li$_5$Si$_2$). Negative values of $E_d$ therefore indicate compounds that break the existing MP convex hull.
Notably, CHGGen successfully predicted Li$_5$Si$_2$ (R-3m, $E_d = -7$ meV/atom) as a thermodynamically stable polymorph beyond the known MP stable structures (Fig.~\ref{fig:LiSi}b). Interestingly, this structure had also been identified through previous studies, including work by \citet{Richard_hennig_Li5Si2} using genetic algorithms and \citet{chris_pickard_Li5Si2} using random structure search coupled with DFT calculations.

Using generative models, one can also identify metastable polymorphs with low decomposition energies in related compositions. For example, structures for compositions corresponding to  Li$_5$Si$_2$ ($E_d = -0.006$ eV/atom), Li$_2$Si ($E_d = 0.003$ eV/atom), and Li$_4$Si ($E_d = 0.010$ eV/atom) with C2/m spacegroup are shown in Fig.~\ref{fig:LiSi}c. 
These low-energy polymorphs with different local structure motifs provide a more detailed mapping of the potential energy landscape, which benefits the training of MLIPs to understand the Si network aggregation and its effect on Li transport kinetics \cite{Sivonxay2020_LiSi}.
The case study illustrates how practitioners can benefit from combining diffusion models and foundation potentials to explore continuous unknown chemical spaces, such as those encountered in alloy design.

\section{Discussion}

Recent advances in diffusion models have shown their promise for crystal structure prediction (CSP), leveraging their ability to learn geometric features beyond elemental substitution heuristics \cite{Park2024_generative}. Graph neural networks have emerged as the preferred architectural choice for diffusion-based approaches, primarily due to their inherent capability to incorporate rotational equivariance \cite{gasteiger_gemnet_2022, Batzner2022_nequip}. 
A particularly significant theoretical insight is that the score function derived from the GNN-based diffusion models is mathematically equivalent to interatomic forces under harmonic potential approximation \cite{xie2021_cdvae}. 
This equivalence reveals that denoising pretraining provides substantial benefits for interatomic potential modeling \cite{zaidi2022_denoise_pretrain}, which enhances the efficiency of local energy minima exploration as evidenced by the leading performance in MatBench Discovery benchmarks \cite{Riebesell2023, liao2024_DeNS}. 

Based on this concept, diffusion models with GNNs represent a logical framework for generating reasonable local structure motifs (e.g., atomic bonding patterns), which proves valuable when optimizing local atomic arrangements in CSP problems.
Nonetheless, their inherent locality bias limits their ability to capture long-range periodic orders. 
\citet{Gong2023} revealed that state-of-the-art GNNs fall short of accurately capturing the periodicity of crystal structures, i.e., lattice parameters. 
This fundamental limitation explains the diminished performance of GNN-based diffusion models when generating structures with large unit cells, where long-period crystallinity is significant but hard to capture. This locality bias is exacerbated for species that can adapt to diverse local environments (e.g., Li), which have a broad distribution of stable coordination geometries. The unconditional generation defaults to producing a disordered "mosaic" of these competing local motifs rather than a coherent crystal structure.

\begin{figure}[tb]
    \centering
    \includegraphics[width=0.95\linewidth]{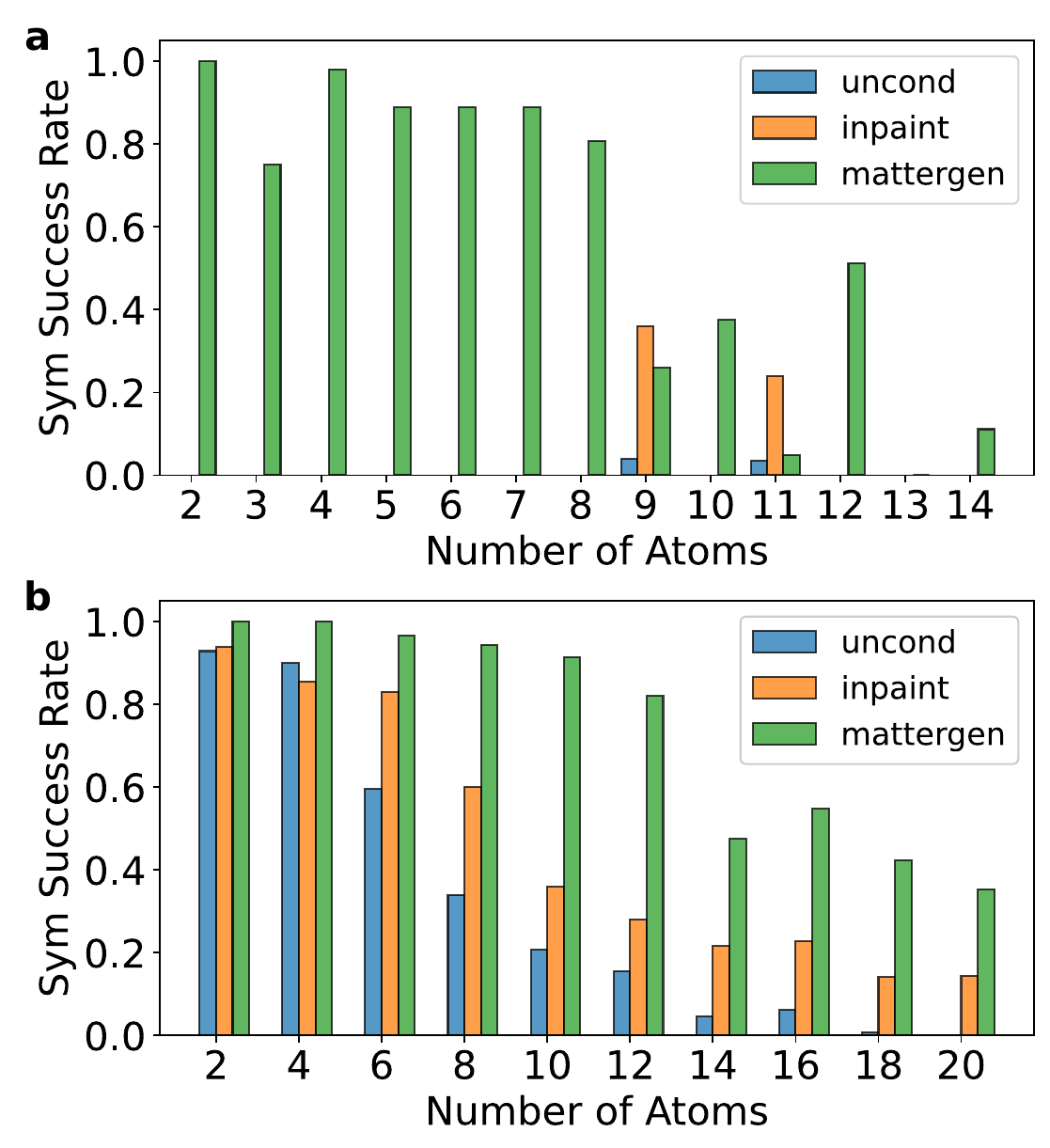}
    \caption{Comparison of success rates for identifying symmetric crystal structures in (a) the Zn-P-S system and (b) the Li-Si system (blue: unconditional generation with CHGGen; orange: inpainting generation with GHGGen; green: MatterGen with chemical-system-guidance generation).}
    \label{fig:success_mattergen}
\end{figure}

The practical advantage of host-structure-guided generation with inpainting is to augment the symmetric structure generation.
To evaluate this, we compared the success rates of obtaining symmetric crystals across different approaches in Zn-P-S and Li-Si chemical spaces: MatterGen (green), CHGGen with unconditional generation (blue), and CHGGen with inpainting generation (orange).
MatterGen demonstrates superior performance compared to our baseline model for Li-Si alloys (Fig.~\ref{fig:success_mattergen}b), highlighting the importance of explicitly modeling lattice diffusion in conjunction with atomic arrangements.
However, for more complex systems such as Zn-P-S that contain polyanions, MatterGen performs less effectively when the number of atoms exceeds 10, with success rates lower than CHGGen with inpainting generation. This highlights the importance of inpainting generation when dealing with chemical systems involving polyanions. Notably, MatterGen exhibits declining success rates as system size increases in both cases, suggesting that it is likely all GNN-based diffusion models face scalability challenges. This scalability constraint potentially limits their ability to predict complex structural prototypes (e.g., NASICON-type frameworks) across various solid-state materials.

Although CHGGen has not yet achieved industry-level performance as MatterGen in stability or general symmetry success rate, it shows clear improvements in generating symmetric crystal structures compared to unconditional generation methods. Assessing crystallographic symmetry in generated structures is critical for CSP, as discovering new structural prototypes is a key step toward materials discovery \cite{szymanski2025establishing}. These prototypes can subsequently guide elemental substitution strategies \cite{merchant2023scaling} or evolutionary searches \cite{gan2025large} for compositional optimization. The inpainting-based generation approach samples from the conditional distribution of unknown structural components within a given framework, yielding more well-defined local atomic arrangements than \textit{de novo} sampling from a fully unconstrained distribution. Importantly, this inpainting strategy is simple to implement, as both conditional and unconditional generation operate within the same unified model---distinguished only by the masking strategy used during the reverse diffusion process. This modular design enables seamless integration with existing foundational generative models and provides a flexible mechanism for enforcing symmetry constraints during structure generation.

In addition, we highlight the practical significance of CHGGen for the structural modification of materials, which often relys on existing database structures. A notable example is the superionic conductor Li$_{0.388}$Ta$_{0.238}$La$_{0.475}$Cl$_3$, discovered through lithiation of the LaCl$_3$-type host structure with additional aliovalent substitution \cite{Yin2023_LaCl3}. CHGGen framework offers a probabilistic approach to such design task that circumvents the need for topological analysis \cite{he2019crystal} or additional inputs such as DFT-derived charge densities \cite{Shen2020_chargeDensity}, which are often computationally prohibitive or not universally applicable. 

Finally, while our framework demonstrates useful augmentation to diffusion-based generative models, the current symmetry refinement approach remains preliminary as it relies on \texttt{spglib} by simply increasing the tolerance threshold. As a proof-of-concept, this method predominantly yields structures with moderate symmetry (e.g., C2, Cm in monoclinic systems), which may limit the discovery of novel structure prototypes (see Fig.~S1 and S2). Looking forward, several promising approaches have emerged for novel framework generation, including symmetry-constrained diffusion \cite{levy2025symmcd} and prototype-based generation using Wyckoff position-based representations \cite{zhu2023wycryst, cao2024_crysformer, kazeev2025wyckoff}. The integration of these advanced symmetry handling strategies with CHGGen could enhance the discovery of crystal structures particularly with intercalation chemistry.

In summary, we present CHGGen as an integrated framework that combines unconditional and inpainting generation with foundation potential optimization for CSP. While challenges in scaling complexity persist, the inpainting method provides a useful approach for generating symmetric crystals and incorporating intercalants into existing database structures. We anticipate broad adoption of this framework as the modular design of the inpainting methodology enables seamless integration with emerging diffusion models, which will ultimately accelerate materials discovery across diverse chemical spaces.

\section*{Acknowledgements} 
This work was supported by the U.S. Department of Energy, Office of Science, Office of Basic Energy Sciences, Materials Sciences and Engineering Division under Contract No. DE-AC0205CH11231 (Materials Project program KC23MP). The computations were supported by the National Energy Research Scientific Computing Center (NERSC) under the GenAI Project and the National Renewable Energy Laboratory (NREL) clusters under silimorphous allocation. P.Z. acknowledges funding support from the BIDMaP Postdoctoral Fellowship. The authors thank Bingqing Cheng, Yifan Chen, and Aaron Kaplan for valuable discussions.

\section*{Methods}
\textbf{Model architecture.}~We used a customized SE(3)-equivariant graph neural network (GNN) from the implementation in the \verb|graphite| library \cite{Hsu2024} based on the \verb|NequIP| and \verb|e3nn| frameworks \cite{Batzner2022_nequip, geiger2022e3nn}. The GNN model directly outputs a vector to represent the noising displacement. The initial embedding generates two attributes $\boldsymbol{h}_{i,x}$ and $\boldsymbol{h}_{i,z}$ transforming the type of the $i$-th atom through embedding layers. 
The interaction blocks update node attributes $\boldsymbol{h}_{i,x}$ by self-interaction and aggregating attributes of neighboring atoms with the weighted tensor product (wTP), whereas $\boldsymbol{h}_{i,z}$ does not change
\begin{equation}
    \boldsymbol{h}_{i,x}^{(l+1)}  = \mathrm{wTP}(\boldsymbol{h}_{i,x}^{(l)}, \boldsymbol{h}_{i,z}^{(l)}),
    \label{eq:self_interact}
\end{equation}
\begin{equation}
    \boldsymbol{h}_{i,x}^{(l+1)} = \frac{1}{Z} \sum_{j \in N(i)} \mathrm{wTP}_{\Vert \boldsymbol{e}_{ij} \Vert} (\boldsymbol{h}_{j,x}^{(l)} , Y(\hat{\boldsymbol{e}}_{ij})).
    \label{eq:conv_interact}
\end{equation}
Eq.~\eqref{eq:self_interact} and Eq.~\eqref{eq:conv_interact} represent self-interaction and convolution, respectively. $Y(\hat{\boldsymbol{e}}_{ij})$ is the spherical harmonic of a normalized vector $\hat{\boldsymbol{e}}_{ij}$ pointing from node $i$ to $j$. $N(i)$ denotes neighbors of node $i$. The final self-interaction layer outputs a single vector $\boldsymbol{s}_\theta$ representing the estimated score
\begin{equation}
    \boldsymbol{h}_{i,x}^{(L)}  = \mathrm{wTP}(\boldsymbol{h}_{i,x}^{(L-1)}, \boldsymbol{h}_{i,z}^{(L-1)}) = (\boldsymbol{s}_\theta)_i
    \label{eq:GNN_score}
\end{equation}
The training framework was built upon the architecture of \texttt{CDVAE} using the \texttt{pytorch-lighting} library \cite{xie2021_cdvae}.

\textbf{Dataset.}~The training dataset in our work consists of 109,805 crystal structures from the Materials Project (MP) database with $E_{\mathrm{hull}} < 0.1$ eV. The $E_{\mathrm{hull}}$ represents the energy above the convex hull in phase diagrams, which provides the thermodynamic stability of a material at 0~K ($E_{\mathrm{hull}} = 0$ for stable materials, and $E_{\mathrm{hull}} > 0$ for meta-stable materials) \cite{Bartel2020_critical}. The training set is divided into training and validation sets at an 8:2 ratio. The test data set consists of 1,131 stable ionic crystal structures ($E_{\mathrm{hull}} = 0$ eV, containing Li/Mg/Na/Ca ions) from the WBM data set \cite{WBM_2021}, which does not have overlapping crystal structures with the MP database. 
The test performance of the model is reported in Ref.~\cite{dai2024inpainting}. 

\textbf{Model evaluation.}~We evaluated the difference
between the generated structures and the relaxed structures using the StructureMatcher from pymatgen \cite{Ong2013_pymatgen} to get the root-
mean-squared distance (RMSD) with the maximum pair-wise distance (max RMSD) as a metric of local stability. 
The success rate is defined as the fraction of structures excluding low-symmetry space groups among all generated structures. We used \{P1, P-1\} to exclude the low-symmetry space groups for the Zn-P-S system, and \{P1, P-1, Pm\} for the Li-Si system.

\textbf{Hyperparameters.}~In the random search for Bravais lattices, the ratio of $c/a$ and $b/a$ is sampled uniformly between $\mathcal{N}(-0.75, 1.25)$. For both the unconditional and inpainting generation, a signal-to-noise ratio $\delta$ of 0.4 is used. The spacegroup matching is performed using tolerances with site tolerance (\texttt{stol}) from 0.1 to 2.0, and angle tolerance from 10 to 30 degrees, by increasing the tolerance to a higher value if the spacegroup matching results in low symmetry. The atomic volume $V_0$ is set to 24 \AA$^3$ for Zn-P-S and 16 \AA$^3$ for Li-Si generations. For the inpainting process, three reverse diffusion generations were run for each refined framework. After the CHGNet/DFT relaxation, a site tolerance of 0.1 and an angle tolerance of 10 degree are used to get the refined crystal structure.

\textbf{Machine learning interatomic potential.}~For the generated structures, we used the pre-trained CHGNet as an efficient and accurate calculator to evaluate the energy and interatomic forces. The pre-trained CHGNet achieves mean-absolute errors of 30 meV/atom for energy and 70 meV/\AA\ for interatomic forces on the MPtrj dataset against the DFT at GGA/GGA$+U$ level of accuracy \cite{Deng2023_chgnet}. The structure relaxations were optimized by the \texttt{FIRE} optimizer \cite{Bitzek2006_FIRE} over the potential energy surface provided by CHGNet, where the atom positions, cell shape, and cell volume were optimized to reach converged interatomic forces of 0.1 eV/\AA \cite{Deng2023_chgnet}. 

\textbf{MP phase diagram.} After CHGNet structure relaxation, the thermodynamic stability is evaluated through the decomposition energy ($E_d$) with respect to the convex hull of known stable phases: $E_d = E_{\mathrm{s}} - \sum_i x_i E_i$, where $E_{\mathrm{s}}$ represents the total energy per atom, $x_i$ denotes the molar fraction of the $i$-th competing phase, and $E_i$ corresponds to its ground-state energy per atom. Comparing with $E_\text{hull}$, the decomposition energy $E_d$ can be negative. The convex hull serves as a fixed reference of structures, and the evaluated structure $s$ is not necessarily part of this hull. A negative decomposition energy ($E_d < 0$) indicates a thermodynamically stable state below the convex hull, while $E_d > 0$ suggests a metastable phase with a driving force for decomposition into more stable compounds. Our primary objective is to identify stable crystal structures where $E_d \leq 0$. We used the patched phase diagram (version 2023-02-07-ppd-mp) from the Materials Project. Since the pretrained CHGNet was based on the corrected energies from MP, the output energy prediction is directly compatible with the MP phase diagram with the  \texttt{MaterialsProject2020Compatibility}~\cite{Wang2021_compatability}.

\textbf{DFT calculations.}~We performed density functional theory (DFT) calculations with the \texttt{VASP} package using the projector-augmented wave method \cite{Kresse1996_VASP, Kresse1999_PAW}, a plane-wave basis set with an energy cutoff of 680 eV and a $k$-spacing of 0.22 for Li-Si (metallic) and 0.44 for Zn-P-S (insulator) systems. The calculations were converged to $10^{-6}$ eV in total energy for electronic loops and 0.02 eV/\AA\ in interatomic forces for ionic loops. The regularized strongly constrained and appropriately normed meta-GGA exchange-correlation functional (r$^2$SCAN) \cite{Sun2015_SCAN, Furness2020_r2SCAN} was used with consistent computational settings as \texttt{MPScanRelaxSet} \cite{Kingsbury2022_r2SCAN_PRM}.

\section*{Availability}
The codebase for CHGGen is available at \href{https://github.com/zhongpc/chggen}{\texttt{https://github.com/zhongpc/chggen}}. 

\section*{Conflicts of interest}
The authors declare no conflicts of interest.

\bibliography{references}
\end{document}